\documentclass[fleqn,twoside]{article} 
 \usepackage{epsf,multicol,ifthen}
  \usepackage{ujp-hep}

\usepackage[cp866nav]{inputenc}
 \usepackage[english,russian]{babel}

\usepackage{amsmath,amstext,amssymb,cite}
 \usepackage[dvips]{graphicx}

\newcommand{\Ds}{\displaystyle}                           

\nazva{Resummation in QCD Fractional Analytic Perturbation Theory}%

\nazvacol{Resummation in QCD FAPT}%

\avtor{Alexander P.~Bakulev}
\avtorcol{A.~P.~Bakulev}%

\inst{Bogoliubov Laboratory of Theoretical Physics}%
\adr{Joint Institute for Nuclear Research, Dubna, 141980 Russia}

\begin{document}
\setcounter{page}{1}%
\maketitl
\begin{multicols}{2}
\anot{We describe the generalization of Analytic Perturbation Theory (APT)
for QCD observables,
initiated by Radyushkin, Krasnikov, Pivovarov, Shirkov and Solovtsov,
to fractional powers of coupling --- Fractional APT (FAPT).
The basic aspects of FAPT is shortly summarized.
We describe how to treat heavy-quark thresholds in FAPT
and then show how
to resum perturbative series in both the one-loop APT and FAPT.
As an application we consider FAPT description
of the Higgs boson decay $H^0\to b\bar{b}$.
The main conclusion is:
To achieve an accuracy of the order of 1\%
it is enough
to take into account up to the third correction.}%

\section{APT and FAPT in QCD}
 \label{sec:APT}
In the standard QCD Perturbation Theory (PT) we know that
the Renormalization Group (RG) equation $da_s[L]/dL = -a_s^2-\ldots$
for the effective coupling $\alpha_s(Q^2)=a_s[L]/\beta_f$
with $L=\ln(Q^2/\Lambda^2)$, $\beta_f=b_0(N_f)/(4\pi)=(11-2N_f/3)/(4\pi)$\footnote{%
   We use notations $f(Q^2)$ and $f[L]$ in order to specify the arguments we mean ---
   squared momentum $Q^2$ or its logarithm $L=\ln(Q^2/\Lambda^2)$,
   that is $f[L]=f(\Lambda^2\cdot e^L)$ and $\Lambda^2$ is usually referred to $N_f=3$ region.}.
Then the one-loop solution generates Landau pole singularity,
$a_s[L] = 1/L$.

Strictly speaking the QCD Analytic Perturbation Theory (APT)
was initiated by N.~N.~Bogolyuov et al. paper of 1959~\cite{BLS60},
where ghost-free effective coupling for QED has been constructed.
Then in 1982 Radyushkin \cite{Rad82} and Krasnikov and Pivovarov \cite{KP82}
using the same dispersion technique
suggested regular (for $s\geq \Lambda^2$) QCD running coupling in Minkowskian region,
the famous $\Ds \frac{1}{\pi}\arctan\frac{\pi}{L}$.
After that in 1995 Jones and Solovtsov discovered
the coupling which appears to be finite for all $s$
and coincides with Radyushkin one for $s\geq\Lambda^2$,
namely $\mathfrak A_1[L]$ in Eq.\ (\ref{eq:U_1}).
Just in the same time
Beneke et al.~\cite{BB95,BBB95},
using the renormalon based approach,
and
Shirkov and Solovtsov \cite{SS},
using the same dispersion approach of~\cite{BLS60},
discovered ghost-free coupling $\mathcal A_1[L]$, Eq.\ (\ref{eq:A_1}),
in Euclidean region.

But Shirkov--Solovtsov approach, named APT,
was more powerful:
in Euclidean domain,
$\displaystyle-q^2=Q^2$, $\displaystyle L=\ln Q^2/\Lambda^2$,
it generates the following set of images for the effective coupling
and its $n$-th powers,
$\displaystyle\left\{{\mathcal A}_n[L]\right\}_{n\in\mathbb{N}}$,
whereas in Minkowskian domain,
$\displaystyle q^2=s$, $\displaystyle L_s=\ln s/\Lambda^2$,
it generates another set,
$\displaystyle\left\{{\mathfrak A}_n[L_s]\right\}_{n\in\mathbb{N}}$
(see also in~\cite{Sim01}).
APT is based on the RG and causality
that guaranties standard perturbative UV asymptotics
and spectral properties.
Power series $\sum_{m}d_m a_s^m[L]$
transforms into non-power series
$\sum_{m}d_m {\mathcal A}_{m}[L]$ in APT.

By the analytization in APT for an observable $f(Q^2)$
we mean the ``K\"allen--Lehman'' representation
\begin{eqnarray}
 \label{eq:An.SD}
  \left[f(Q^2)\right]_\text{an}
   = \int_0^{\infty}\!
      \frac{\rho_f(\sigma)}
         {\sigma+Q^2-i\epsilon}\,
       d\sigma
\end{eqnarray}
with $\Ds\rho_f(\sigma)=\frac{1}{\pi}\,\textbf{Im}\,\big[f(-\sigma)\big]$.
Then in the one-loop approximation $\rho_1(\sigma)=1/\sqrt{L_\sigma^2+\pi^2}$ and
\begin{subequations}
 \label{eq:A.U}
 \begin{eqnarray}
  \label{eq:A_1}
 \mathcal A_1[L]
  \!\!&\!\!=\!\!&\!\! \int_0^{\infty}\!\frac{\rho_1(\sigma)}{\sigma+Q^2}\,d\sigma\
   =\ \frac{1}{L} - \frac{1}{e^L-1}\,,~\\
 \label{eq:U_1}
 {\mathfrak A}_1[L_s]
  \!\!&\!\!=\!\!&\!\! \int_s^{\infty}\!\frac{\rho_1(\sigma)}{\sigma}\,d\sigma\
   =\ \frac{1}{\pi}\,\arccos\frac{L_s}{\sqrt{\pi^2+L_s^2}}\,,~
 \end{eqnarray}
\end{subequations}
whereas analytic images of the higher powers ($n\geq2, n\in\mathbb{N}$) are:
\begin{eqnarray}
 \label{eq:recurrence}
 {\mathcal A_n[L] \choose \mathfrak A_n[L_s]}
  \!\!&\!\!=\!\!&\!\! \frac{1}{(n-1)!}\left( -\frac{d}{d L}\right)^{n-1}
      {\mathcal A_{1}[L] \choose \mathfrak A_{1}[L_s]}\,.
\end{eqnarray}

At first glance, the APT is a complete theory
providing tools to produce
an analytic answer for any perturbative series in QCD.
But in 2001 Karanikas and Stefanis~\cite{KS01}
suggested the principle of analytization ``as a whole''
in the $Q^2$ plane for hadronic observables,
calculated perturbatively.
More precisely, they proposed the analytization recipe
for terms like
$\int_{0}^{1}\!dx\!\int_{0}^{1}\!dy\,
  \alpha_\text{s}\left(Q^{2}xy\right) f(x)f(y)$,
which can be treated as an effective account
for the logarithmic terms
in the next-to-leading-order
approximation of the perturbative QCD.
This actually generalizes the analytic approach
suggested in~\cite{SSK9900}.
Indeed, in the standard QCD PT one has also:\\
(i) the factorization procedure in QCD
    that gives rise to the appearance of logarithmic factors of the type:
     $a_s^\nu[L]\,L$;\\
(ii) the RG evolution
     that generates evolution factors of the type:
     $B(Q^2)=\left[Z(Q^2)/Z(\mu^2)\right]$ $B(\mu^2)$,
     which reduce in the one-loop approximation to
     $Z(Q^2) \sim a_s^\nu[L]$ with $\nu=\gamma_0/(2b_0)$
     being a fractional number.\\
All that means
that in order to generalize APT
in the ``analytization as a whole'' direction
one needs to construct analytic images
of new functions:
$\displaystyle a_s^\nu,~a_s^\nu\,L^m, \ldots$\,.
This task has been performed in the frames of the so-called FAPT,
suggested in~\cite{BMS-APT,BKS05}.
Now we briefly describe this approach.

In the one-loop approximation
using recursive relation (\ref{eq:recurrence})
we can obtain explicit expressions for
${\mathcal A}_{\nu}[L]$
and ${\mathfrak A}_{\nu}[L]$:
\begin{subequations}
\begin{eqnarray}
 {\mathcal A}_{\nu}[L]
 \!\!&\!\!=\!\!&\!\!
   \frac{1}{L^\nu}
  - \frac{F(e^{-L},1-\nu)}{\Gamma(\nu)}\,;
 ~
 \\
 {\mathfrak A}_{\nu}[L]
 \!\!&\!\!=\!\!&\!\!
   \frac{\text{sin}\left[(\nu -1)\arccos\left(\frac{L}{\sqrt{\pi^2+L^2}}\right)\right]}
         {\pi(\nu -1) \left(\pi^2+L^2\right)^{(\nu-1)/2}}\,.~
\end{eqnarray}
\end{subequations}
Here $F(z,\nu)$ is reduced Lerch transcendental function,
which is an analytic function in $\nu$.
They have very interesting properties,
which we discussed extensively in our previous papers~\cite{BMS-APT,BKS05,AB08,Ste09}.

Construction of FAPT with fixed number of quark flavors, $N_f$,
is a two-step procedure:
we start with the perturbative result $\left[a_s(Q^2)\right]^{\nu}$,
generate the spectral density $\rho_{\nu}(\sigma)$ using Eq.\ (\ref{eq:An.SD}),
and then obtain analytic couplings
${\mathcal A}_{\nu}[L]$ and ${\mathfrak A}_{\nu}[L]$ via Eqs.\ (\ref{eq:A.U}).
Here $N_f$ is fixed and factorized out.
We can proceed in the same manner for $N_f$-dependent quantities:
$\left[\alpha_s^{}(Q^2;N_f)\right]^{\nu}$
$\Rightarrow$
$\bar{\rho}_{\nu}(\sigma;N_f)=\bar{\rho}_{\nu}[L_\sigma;N_f]
 \equiv\rho_{\nu}(\sigma)/\beta_f^{\nu}$
$\Rightarrow$
$\bar{\mathcal A}_{\nu}^{}[L;N_f]$ and $\bar{\mathfrak A}_{\nu}^{}[L;N_f]$ ---
here $N_f$ is fixed, but not factorized out.

Global version of FAPT~\cite{AB08},
which takes into account heavy-quark thresholds,
is constructed along the same lines
but starting from global perturbative coupling
$\left[\alpha_s^{\,\text{\tiny glob}}(Q^2)\right]^{\nu}$,
being a continuous function of $Q^2$
due to choosing different values of QCD scales $\Lambda_f$,
corresponding to different values of $N_f$.
We illustrate here the case of only one heavy-quark threshold
at $s=m_4^2$,
corresponding to the transition $N_f=3\to N_f=4$.
Then we obtain the discontinuous spectral density
\begin{eqnarray}
 \rho_n^\text{\tiny glob}(\sigma)
  \!\!&\!\!=\!\!&\!\! \theta\left(L_\sigma<L_{4}\right)\,
       \bar{\rho}_n\left[L_\sigma;3\right]
  \nonumber\\
  \!\!&\!\!+\!\!&\!\! \theta\left(L_{4}\leq L_\sigma\right)\,
       \bar{\rho}_n\left[L_\sigma+\lambda_4;4\right]\,,~
 \label{eq:global_PT_Rho_4}
\end{eqnarray}
with $L_{\sigma}\equiv\ln\left(\sigma/\Lambda_3^2\right)$,
$L_{f}\equiv\ln\left(m_f^2/\Lambda_3^2\right)$
and
$\lambda_f\equiv\ln\left(\Lambda_3^2/\Lambda_f^2\right)$ for $f=4$,
which is expressed in terms of fixed-flavor spectral densities
with 3 and 4 flavors,
$\bar{\rho}_n[L;3]$ and $\bar{\rho}_n[L+\lambda_4;4]$.
However it generates the continuous Minkowskian coupling
\begin{subequations}
\begin{eqnarray}
 {\mathfrak A}_{\nu}^{\text{\tiny glob}}[L]
  \!\!&\!\!=\!\!&\!\!
    \theta\left(L\!<\!L_4\right)
     \Bigl(\bar{{\mathfrak A}}_{\nu}^{}[L;3]
          + \Delta_{43}\bar{{\mathfrak A}}_{\nu}^{}
     \Bigr)
  \nonumber\\
  \!\!&\!\!+\!\!&\!\!
    \theta\left(L_4\!\leq\!L\right)\,
     \bar{{\mathfrak A}}_{\nu}^{}[L+\lambda_4;4]\,.
 \label{eq:An.U_nu.Glo.Expl}
\end{eqnarray}
with $\Delta_{43}\bar{{\mathfrak A}}_{\nu}^{}=
            \bar{{\mathfrak A}}_{\nu}^{}[L_4+\lambda_4;4]
          - \bar{{\mathfrak A}}_{\nu}^{}[L_4;3]
$
and the analytic Euclidean coupling ${\cal A}_{\nu}^{\text{\tiny glob}}[L]$
\begin{eqnarray}
 {\cal A}_{\nu}^{\text{\tiny glob}}[L]
  \!\!&\!\!=\!\!&\!\! \bar{{\cal A}}_{\nu}^{}[L+\lambda_4;4]
 \nonumber\\
  \!\!&\!\!+\!\!&\!\! \int\limits_{-\infty}^{L_4}\!
       \frac{\bar{\rho}_{\nu}^{}[L_\sigma;3]
            -\bar{\rho}_{\nu}^{}[L_\sigma+\lambda_{4};4]}
            {1+e^{L-L_\sigma}}\,
         dL_\sigma
  \label{eq:Delta_f.A_nu}
\end{eqnarray}
\end{subequations}
(for more detail see in~\cite{AB08}).

\section{Resummation in the one-loop APT and FAPT}
\label{sec:Resum.FAPT}
We consider now the perturbative expansion
of a typical physical quantity,
like the Adler function and the ratio $R$,
in the one-loop APT.
Due to limited space of our presentation
we provide all formulas only
for quantities in Minkowski region:
\begin{eqnarray}
 \label{eq:APT.Series}
  \mathcal R[L]
   = \sum_{n=1}^{\infty}
      d_n\,\mathfrak A_{n}[L]\,.
\end{eqnarray}
We suggest that there exist the generating function $P(t)$
for coefficients $\tilde{d}_n=d_n/d_1$:
\begin{equation}
 \tilde{d}_n
  =\int_{0}^\infty\!\!P(t)\,t^{n-1}dt
   ~~~\text{with}~~~
   \int_{0}^\infty\!\!P(t)\,d t = 1\,.
 \label{eq:generator}
\end{equation}
To shorten our formulae, we use for the integral
$\int_{0}^{\infty}\!\!f(t)P(t)dt$
the following notation:
$\langle\langle{f(t)}\rangle\rangle_{P(t)}$.
Then coefficients $d_n = d_1\,\langle\langle{t^{n-1}}\rangle\rangle_{P(t)}$
and as has been shown in~\cite{MS04}
we have the exact result for the sum in (\ref{eq:APT.Series})
\begin{eqnarray}
 \label{eq:APT.Sum.DR[L]}
  \mathcal R[L]
   = d_1\,\langle\langle{\mathfrak A_1[L-t]}\rangle\rangle_{P(t)}\,.
\end{eqnarray}
The integral in variable $t$ here has a rigorous meaning,
ensured by the finiteness of the coupling  $\mathfrak A_1[t] \leq 1$
and fast fall-off of the generating function $P(t)$.

In our previous publications~\cite{AB08,BM08}
we have constructed generalizations of these results,
first, to the case of the global APT,
when heavy-quark thresholds are taken into account.
Then one starts with the series
of the type (\ref{eq:APT.Series}),
where $\mathfrak A_{n}[L]$
are substituted by their global analogs
$\mathfrak A_{n}^\text{\tiny glob}[L]$
(note that due to different normalizations of global
 couplings, $\mathfrak A_{n}^\text{\tiny glob}[L]\simeq\mathfrak A_{n}[L]/\beta_f$,
 the coefficients $d_n$ should be also changed).
Then
\begin{eqnarray}
 \mathcal R^\text{\tiny glob}[L]
  \!\!&\!\!=\!\!&\!\!  d_1 \theta(L\!<\!L_4)
          \langle\langle{
           \Delta_{4}\bar{\mathfrak A}_{1}[t]
           + \bar{\mathfrak A}_{1}\!\Big[L\!-\!\frac{t}{\beta_3};3\Big]
           }\rangle\rangle_{P(t)}\nonumber\\
 \!\!\!&\!\!+\!\!&\!\!\! d_1  \theta(L\!\geq\!L_4)
          \langle\langle{
           \bar{\mathfrak A}_{1}\!\Big[L\!+\!\lambda_4-\!\frac{t}{\beta_4};4\Big]
           }\rangle\rangle_{P(t)}\,;~~~
 \label{eq:sum.R.Glo.4}
\end{eqnarray}
where $\Delta_4\bar{\mathfrak A}_\nu[t]\equiv
  \bar{\mathfrak A}_\nu\!\Big[L_4+\lambda_{4}-t/\beta_4;4\Big]
 -\bar{\mathfrak A}_\nu\!\Big[L_3-t/\beta_3;3\Big]$.

The second generalization has been obtained for the case
of the global FAPT.
Then the starting point is the series of the type
$\sum_{n=0}^{\infty} d_n\,\mathfrak A_{n+\nu}^\text{\tiny glob}[L]$
and the result of summation is a complete analog of Eq.\ (\ref{eq:sum.R.Glo.4})
with substitutions
\begin{eqnarray}
 \label{eq:P_nu(t)}
  P(t)\Rightarrow P_{\nu}(t) =
   \int_0^{1}\!P\left(\frac{t}{1-x}\right)
    \frac{\nu\,x^{\nu-1}dx}
         {1-x}\,,
\end{eqnarray}
$d_0\Rightarrow d_0\,\bar{\mathfrak A}_{\nu}[L]$,
$\bar{\mathfrak A}_{1}[L-t]\Rightarrow
 \bar{\mathfrak A}_{1+\nu}[L-t]$,
and
$\Delta_4\bar{\mathfrak A}_{1}[t]\Rightarrow
 \Delta_4\bar{\mathfrak A}_{1+\nu}[t]$.
All needed formulas have been also obtained
in parallel for the Euclidean case.

\section{Applications to Higgs boson decay}
\label{sec:Appl.Higgs}
Here we analyze the Higgs boson decay to a $\bar{b}b$ pair.
For its width we have
\begin{eqnarray}
 \label{eq:Higgs.decay.rate}
  \Gamma(\text{H} \to b\bar{b})
  = \frac{G_F}{4\sqrt{2}\pi}\,
     M_{H}\,
      \widetilde{R}_\text{\tiny S}(M_{H}^2)
\end{eqnarray}
with $\widetilde{R}_\text{\tiny S}(M_{H}^2)
     \equiv m^2_{b}(M_{H}^2)\,R_\text{\tiny S}(M_{H}^2)$
and
$R_\text{\tiny S}(s)$
is the $R$-ratio for the scalar correlator,
see for details in~\cite{BMS-APT,BCK05}.
In the one-loop FAPT this generates the following
non-power expansion\footnote{%
Appearance of denominators $\pi^n$ in association
with the coefficients $\tilde{d}_n$
is due to $d_n$ normalization.}:
\begin{eqnarray}
 \widetilde{\mathcal R}_\text{\tiny S}[L]
   =  3\,\hat{m}_{(1)}^2\,
      \Bigg\{\mathfrak A_{\nu_{0}}^{\text{\tiny glob}}[L]
          + d_1^\text{\,\tiny S}\,\sum_{n\geq1}
             \frac{\tilde{d}_{n}^\text{\,\tiny S}}{\pi^{n}}\,
              \mathfrak A_{n+\nu_{0}}^{\text{\tiny glob}}[L]
      \Bigg\}\,,
 \label{eq:R_S-MFAPT}
\end{eqnarray}
where $\hat{m}_{(1)}^2=9.05\pm0.09$~GeV$^2$ is the RG-invariant
of the one-loop $m^2_{b}(\mu^2)$ evolution
$m_{b}^2(Q^2) = \hat{m}_{(1)}^2\,\alpha_{s}^{\nu_{0}}(Q^2)$
with $\nu_{0}=2\gamma_0/b_0(5)=1.04$ and
$\gamma_0$ is the quark-mass anomalous dimension.
This value $\hat{m}_{(1)}^2$ has been obtained
using the one-loop relation~\cite{KK08}
between the pole $b$-quark mass of~\cite{KuSt01}
and the mass $m_b(m_b)$.

We take for the generating function $P(t)$
the Lipatov-like model of~\cite{BM08}
with $\left\{c=2.4,~\beta=-0.52\right\}$
\begin{subequations}
\label{eq:Higgs.Model}
\begin{eqnarray}
  \tilde{d}_{n}^\text{\,\tiny S}
   \!\!&\!\!=\!\!&\!\! c^{n-1}\frac{\Gamma (n+1)+\beta\,\Gamma (n)}{1+\beta}\,,
 \\
  P_\text{\tiny S}(t)
   \!\!&\!\!=\!\!&\!\! \frac{(t/c)+\beta}{c\,(1+\beta)}\,e^{-{t/c}}\,.
\end{eqnarray}
\end{subequations}
It gives a very good prediction for
$\tilde{d}_{n}^\text{\,\tiny S}$ with $n=2, 3, 4$,
calculated in the QCD PT~\cite{BCK05}:
$7.50$, $61.1$, and  $625$
in comparison with
$7.42$, $62.3$, and  $620$.
Then we apply FAPT resummation technique
to estimate
how good is FAPT
in approximating the whole sum $\widetilde{\mathcal R}_\text{\tiny S}[L]$
in the range $L\in[11.5,13.7]$
which corresponds to the range
$M_H\in[60,180]$~GeV$^2$
with $\Lambda^{N_f=3}_{\text{QCD}}=189$~MeV
and ${\mathfrak A}^{\text{\tiny glob}}_{1}(m_Z^2)=0.122$.
In this range we have ($L_6=\ln(m_t^2/\Lambda_3^2)$)
\begin{eqnarray}
 \frac{\widetilde{\mathcal R}_\text{\tiny S}[L]}
      {3\,\hat{m}_{(1)}^2}
  \!\!&\!\!=\!\!&\!\! {\mathfrak A}^\text{\tiny glob}_{\nu_{0}}[L]
   + \frac{d_{1}^\text{\,\tiny S}}{\pi}\,
      \langle\langle{\bar{\mathfrak A}_{1+\nu_{0}}\!
                      \Big[L\!+\!\lambda_5\!-\!\frac{t}{\pi\beta_5};5\Big]
      }\rangle\rangle_{P_{\nu_{0}}^\text{\,\tiny S}}
  \nonumber\\
  \!\!&\!\!\!\!&~~~~~~~~~\,
   +\,\frac{d_{1}^\text{\,\tiny S}}{\pi}\,
      \langle\langle{\Delta_{6}\bar{\mathfrak A}_{1+\nu_{0}}
                      \left[\frac{t}{\pi}\right]
      }\rangle\rangle_{P_{\nu_{0}}^\text{\,\tiny S}}~
 \label{eq:R_S.Sum}
\end{eqnarray}
with $P_{\nu_{0}}^\text{\,\tiny S}(t)$ defined via Eqs.\ (\ref{eq:Higgs.Model})
and (\ref{eq:P_nu(t)}).

Now we analyze the accuracy of the truncated FAPT expressions
\begin{eqnarray}
 \label{eq:FAPT.trunc}
 \widetilde{\mathcal R}_\text{\tiny S}[L;N]
  \!\!&\!\!=\!\!&\!\! 3\,\hat{m}_{(1)}^2\,
       \left[{\mathfrak A}_{\nu_{0}}^{\text{\tiny glob}}[L]
           + d_1^\text{\,\tiny S}\,\sum_{n=1}^{N}
              \frac{\tilde{d}_{n}^\text{\,\tiny S}}{\pi^{n}}\,
               {\mathfrak A}_{n+\nu_{0}}^{\text{\tiny glob}}[L]
       \right]
\end{eqnarray}
and compare them with the total sum
$\widetilde{\mathcal R}_\text{\tiny S}[L]$
in Eq.\ (\ref{eq:R_S.Sum})
using relative errors
$\Delta_N[L]=1-\widetilde{\mathcal R}_\text{\tiny S}[L;N]/\widetilde{\mathcal R}_\text{\tiny S}[L]$.
In Fig.~1
we show these errors for $N=2$, $N=3$, and $N=4$
in the analyzed range of $L\in[11,13.8]$.
We see that already $\widetilde{\mathcal R}_\text{\tiny S}[L;2]$
gives accuracy of the order of 2.5\%,
whereas $\widetilde{\mathcal R}_\text{\tiny S}[L;3]$
of the order of 1\%.
\begin{center}
\noindent \epsfxsize=\columnwidth\epsffile{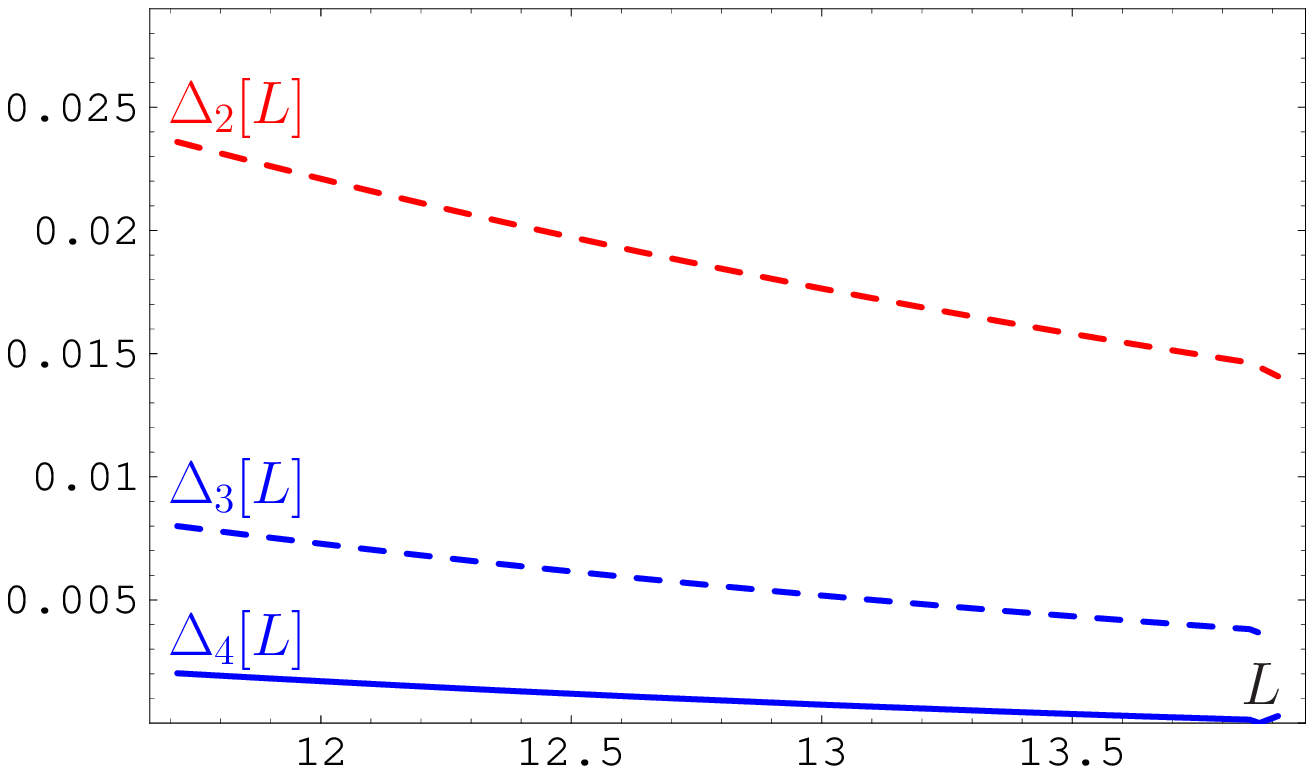}
\end{center}

\vskip-3mm\noindent{\footnotesize Fig.~1. The relative errors
 $\Delta_N[L]$, $N=2, 3$ and $4$, of the truncated FAPT
 in comparison with the exact summation result,
 Eq.\ (\ref{eq:R_S.Sum}).}%
\vskip15pt
Looking in Fig.\ 1 we understand that only in order to have the accuracy better
than 0.5\% we need to take into account the 4-th correction.
We verified also that the uncertainty due to $P(t)$-modelling is small $\lesssim0.6\%$,
while the on-shell mass uncertainty is of the order of $2\%$.
The overall uncertainty then is of the order of $3\%$,
see in Fig.\ 2.

\begin{center}
\noindent \epsfxsize=\columnwidth\epsffile{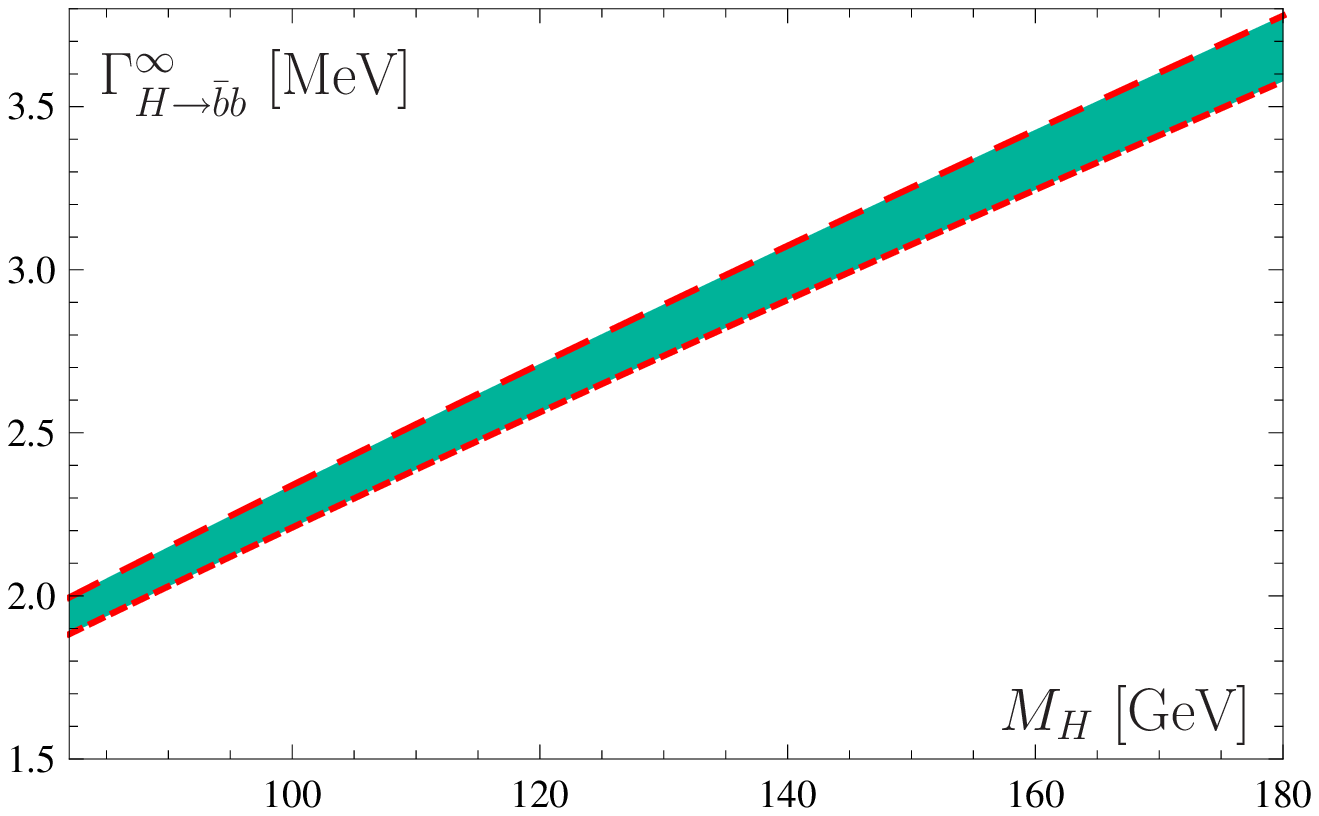}
\end{center}

\vskip-3mm\noindent{\footnotesize Fig.~2. The width $\Gamma_{H\to b\bar{b}}$
 as a function of the  Higgs boson mass $M_{H}$ in the resummed FAPT.
 The width of the shaded strip is due to the overall uncertainties,
 induced by the uncertainties of the resummation procedure and
 the pole mass error-bars.}%
\vskip15pt

\section{Conclusions}
\label{sec:Concl}
In this report we described the resummation approach
in the global versions of the one-loop APT and FAPT
and argued that it produces finite answers,
provided the generating function $P(t)$
of perturbative coefficients $d_n$ is known.
The main conclusion is:
To achieve an accuracy of the order of 1\%
it is enough
to take into account up to the third correction---in complete agreement
with Kataev\&Kim~\cite{KK08}.
The $d_4$ coefficient value is needed only to estimate
corresponding generating functions $P(t)$.

\section*{Acknowledgements}
This work was supported in part by
the Russian Foundation for Fundamental Research,
grants No.\ №~07-02-91557 and 08-01-00686,
the BRFBR--JINR Cooperation Programme,
contract No.\ F08D-001, and
the Heisenberg--Landau Programme under grant 2009.


\end{multicols}

\end{document}